\documentclass[sigconf]{acmart}

\usepackage{epsfig, epstopdf, graphicx, balance}
\usepackage{amsmath,amssymb,mathrsfs,amsfonts}
\usepackage{microtype} 
\usepackage{tabulary} 
\usepackage{threeparttable} 
\usepackage{hhline}

\ifdefined\IEEEtran\else
\usepackage{enumitem}
\setlist[itemize]{itemsep=3pt,topsep=0pt,parsep=0pt,partopsep=3pt,leftmargin=1em}
\setlist[enumerate]{itemsep=3pt,topsep=0pt,parsep=0pt,partopsep=3pt,leftmargin=1em}
\fi

\usepackage{hyperref}

\usepackage[T1]{fontenc}

\usepackage{amsthm}
\theoremstyle{definition}
\newtheorem{defn}{Definition} 
\theoremstyle{plain}

\usepackage{mathtools}

\usepackage{interval}
\intervalconfig{soft open fences}

\usepackage[ruled,vlined]{algorithm2e}
\SetAlgorithmName{Algorithm}{Algorithm}{Algorithm}
\newcommand{\AlgoFontSize}{\small}
\SetCommentSty{textnormal}
\SetNlSty{}{}{:}
\SetAlgoNlRelativeSize{0}
\SetKwInput{KwGlobal}{Global}
\SetKwInput{KwPrecondition}{Precondition}
\SetKwProg{Proc}{Procedure}{:}{}
\SetKwProg{Func}{Function}{:}{}
\SetKw{And}{and}
\SetKw{Or}{or}
\SetKw{To}{to}
\SetKw{DownTo}{downto}
\SetKw{Break}{break}
\SetKw{Continue}{continue}
\SetKw{SuchThat}{\textit{s.t.}}
\SetKw{WithRespectTo}{\textit{wrt}}
\SetKw{Iff}{\textit{iff.}}
\SetKw{MaxOf}{\textit{max of}}
\SetKw{MinOf}{\textit{min of}}
\SetKwBlock{Match}{match}{}{}

\newcommand*{\cTrue}{\textnormal{\tt\it true}}

\usepackage{array}
\newcolumntype{L}[1]{>{\raggedright\let\newline\\\arraybackslash\hspace{0pt}}m{#1}}
\newcolumntype{C}[1]{>{\centering\let\newline\\\arraybackslash\hspace{0pt}}m{#1}}
\newcolumntype{R}[1]{>{\raggedleft\let\newline\\\arraybackslash\hspace{0pt}}m{#1}}

\usepackage{color}
\usepackage{soul}
\soulregister\cite7
\soulregister\ref7
\soulregister\pageref7
\soulregister\autoref7
\soulregister\eqref7

\ifdefined\NoHighlight

\fi

\renewcommand{\eqref}{Equation~\ref}

\newcommand*{\PicDir}{.}
\newcommand*{\ExpDir}{.}
\newcommand*{\Pic}[2]{\PicDir /#2.#1}
\newcommand*{\Exp}[2]{\ExpDir /#2.#1}

\newcommand{\todo}[1]{\hlyellow{\textbf{[}#1\textbf{]}}}
\renewcommand{\todo}[1]{#1}

\newcommand*{\dagor}{DAGOR}
\newcommand*{\codel}{CoDel}
\newcommand*{\seda}{SEDA}

\newcommand*{\wechat}{WeChat}
\newcommand*{\moments}{Moments}
\newcommand*{\wechatpay}{\wechat{} Pay}

\newcommand*{\vB}{\mathcal{B}} 
\newcommand*{\vBmin}{\vB_{L}} 
\newcommand*{\vBmax}{\vB_{H}} 
\newcommand*{\vBopt}{\vB^{\ast}} 
\newcommand*{\vU}{\mathcal{U}} 
\newcommand*{\vUmin}{\vU_{L}} 
\newcommand*{\vUmax}{\vU_{H}} 
\newcommand*{\vUopt}{\vU^{\ast}} 
\newcommand*{\vN}{N} 
\newcommand*{\vNadm}{\vN_{\textnormal{adm}}} 
\newcommand*{\vNexp}{\vN_{\textnormal{exp}}} 
\newcommand*{\vNprefix}{\vN_{\textnormal{prefix}}} 
\newcommand*{\vFoverload}{f_{\textnormal{ol}}} 

\SetKwFunction{fResetHistogram}{ResetHistogram}
\SetKwFunction{fUpdateHistogram}{UpdateHistogram}
\SetKwFunction{fUpdateAdmitLevel}{UpdateAdmitLevel}
\SetKwFunction{fAdmitRequest}{AdmitRequest}

\newcommand*{\AMX}{$\mathcal{M}^{x}$}
\newcommand*{\AM}{$\mathcal{M}^{1}$}
\newcommand*{\AMM}{$\mathcal{M}^{2}$}
\newcommand*{\AMMM}{$\mathcal{M}^{3}$}
\newcommand*{\AMMMM}{$\mathcal{M}^{4}$}

\soulregister\wechat7
\soulregister\moments7
\soulregister\wechatpay7

\acmYear{2018}\copyrightyear{2018}
\setcopyright{acmcopyright}
\acmConference[SoCC '18]{SoCC '18: ACM Symposium on Cloud Computing}{October 11--13, 2018}{Carlsbad, CA, USA}
\acmBooktitle{SoCC '18: ACM Symposium on Cloud Computing, October 11--13, 2018, Carlsbad, CA, USA}
\acmPrice{15.00}
\acmDOI{10.1145/3267809.3267823}
\acmISBN{978-1-4503-6011-1/18/10}

\begin{document}

\title{Overload Control for Scaling \wechat{} Microservices}
\subtitle{(Updated on 18 December 2018)}
\subtitlenote{Merged the errata published in \url{https://arxiv.org/abs/1806.04075v2}.}

\author{Hao Zhou}
\affiliation{%
  \institution{Tencent Inc.}
  \country{China}}
\email{harveyzhou@tencent.com}

\author{Ming Chen}
\affiliation{%
  \institution{Tencent Inc.}
  \country{China}}
\email{mingchen@tencent.com}

\author{Qian Lin}
\affiliation{%
  \institution{National University of Singapore}
  \country{Singapore}}
\email{linqian@comp.nus.edu.sg}

\author{Yong Wang}
\affiliation{%
  \institution{Tencent Inc.}
  \country{China}}
\email{darwinwang@tencent.com}

\author{Xiaobin She}
\affiliation{%
  \institution{Tencent Inc.}
  \country{China}}
\email{stevenshe@tencent.com}

\author{Sifan Liu}
\affiliation{%
  \institution{Tencent Inc.}
  \country{China}}
\email{stephenliu@tencent.com}

\author{Rui Gu}
\affiliation{%
  \institution{Columbia University}
  \state{New York}
  \country{USA}}
\email{ruigu@cs.columbia.edu}

\author{Beng Chin Ooi}
\affiliation{%
  \institution{National University of Singapore}
  \country{Singapore}}
\email{ooibc@comp.nus.edu.sg}

\author{Junfeng Yang}
\affiliation{%
  \institution{Columbia University}
  \state{New York}
  \country{USA}}
\email{junfeng@cs.columbia.edu}

\renewcommand{\shortauthors}{H.~Zhou et~al.}

\begin{abstract}

Effective overload control for large-scale online service system is crucial for protecting the system backend from overload.
Conventionally, the design of overload control is ad-hoc for individual service. 
However, service-specific overload control could be detrimental to the overall system due to intricate service dependencies or flawed implementation of service.
Service developers usually have difficulty to accurately estimate the dynamics of actual workload during the development of service. 
Therefore, it is essential to decouple the overload control from service logic. 
In this paper, we propose \dagor{}, an overload control scheme designed for the account-oriented microservice architecture. 
\dagor{} is service agnostic and system-centric. 
It manages overload at the microservice granule such that each microservice monitors its load status in real time and triggers load shedding in a collaborative manner among its relevant services when overload is detected. 
\dagor{} has been used in the \wechat{} backend for five years. 
Experimental results show that \dagor{} can benefit high success rate of service even when the system is experiencing overload, while ensuring fairness in the overload control. 

\end{abstract}



\maketitle

\section{Introduction}
\label{sec:intro}

Overload control aims to mitigate service irresponsiveness when system is experiencing overload.
This is essential for large-scale online applications that needs to enforce 24$\times$7 service availability, despite any unpredictable load surge. 
Although cloud computing facilitates on-demand provisioning, it still cannot solve the problem of overload---service providers are restricted by the computing resources they can afford from the cloud providers, and therefore cloud providers need overload control for the cloud services they provide.

Traditional overload control for simple service architecture presumes a small number of service components with trivial dependencies.
For a stand-alone service, overload control is primarily targeted at the operating system, service runtime and applications~\cite{atc01:Voigt, osdi99:Banga, tocs97:Mogul}.  
For simple multi-tier services, a gateway at the service entry point monitors the load status of the whole system and rejects client requests when necessary to prevent overloading, i.e., load shedding~\cite{icsoc09:Meulenhoff, tc02:Cherkasova, infocom02:Chen}.

However, modern online services become increasingly complex in the architecture and dependency, far beyond what traditional overload control was designed for. 
Modern online services usually adopt the \textit{service-oriented architecture} (SOA)~\cite{Book05:Erl}, which divides the conventional monolithic service architecture into sub-services connected via network protocols. 
\textit{Microservice} architecture, as a specialization of SOA, often comprises hundreds to thousands of sub-services, namely microservices, to support sophisticated applications~\cite{sosp07:DeCandia, Blog15:Mauro}. 
Each microservice runs with a set of processes on one or multiple machines, and communicates with other microservices through message passing.
By decoupling the implementation and deployment of different microservices, the microservice architecture facilitates independent development and update for each microservice, regardless of the underlying programming language and framework.
This yields the flexibility and productivity for cross-team development of complex online applications.

Overload control for large-scale microservice system must cope with the complexity and high dynamics of the system, which could be very challenging in real practice.
First, all the microservices must be monitored.
If any microservice is out of the scope of monitoring, potential overload may emerge at that spot and further ripple through the related upstream microservices. 
As a consequence, system may suffer from cascading overload and eventually get hung, leading to high delay of the affected services.
Nevertheless, it is extremely difficult to rely on some designated microservices or machine quorum to monitor load status, since no microservice or machine owns the view of the fast evolving service deployment. 

Second, it can be problematic to let microservices handle overload independently, due to the complexity of service dependency. 
For example, suppose the processing of a client request relies on $K$ microservices, but all the required microservices are currently overloaded and each of them rejects incoming requests independently with a probability $p$. 
The expectation of the complete processing of a client request is $(1-p)^{K}$.
If $p$ is close to $1$ or $K$ is large, the system throughput tends to vanish under such circumstance.
However, system overloading is not mitigated by the shed workload, since partial processing of the failed requests still consumes the computational resources.
This causes the system to transit into a non-progressive status, which situation is hard to escape.

Third, overload control needs to adapt to the service changes, workload dynamics and external environments.
If each microservice enforces a service-level agreement (SLA) for its upstream services, it would drastically slow down the update progress of this microservice as well as its downstream services, defeating the key advantage of the microservice architecture. 
Similarly, if the microservices have to exchange tons of messages to manage overload in a cooperative manner, they may not be able to adapt to the load surge, while the overload control messages may get discarded due to system overload and even further deteriorate the system overload.

To address the above challenges, we propose an overload control scheme, called \dagor{}, for a large-scale, account-oriented microservice architecture.
The overall mechanism of \dagor{} works as follows. 
When a client request arrives at an entry service, it is assigned with a business priority and a user priority such that all its subsequent triggered microservice requests are enforced to be consistently admitted or rejected with respect to the same priorities.  
Each microservice maintains its own priority thresholds for admitting requests, and monitors its own load status by checking the system-level resource indicator such as the average waiting time of requests in the pending queue.
Once overload is detected in a microservice, the microservice adjusts its load shedding thresholds using an adaptive algorithm that \todo{attempts to shed half of the load}.
Meanwhile, the microservice also informs its immediate upstream microservices about the threshold changes so that client requests can be rejected in the early stage of the microservice pipeline.

\dagor{} overload control employs only a small set of thresholds and marginal coordination among microservices. 
Such lightweight mechanism contributes to the effectiveness and efficiency of overload handling. 
\dagor{} is also service agnostic since it does not require any service-specific information to conduct overload control.
For instance, \dagor{} has been deployed in the \wechat{} business system to cater overload control for all microservices, in spite of the diversity of business logic.
Moreover, \dagor{} is adaptive with respect to service changes, workload dynamics and external environments, making it friendly to the fast evolving microservice system.

While the problem of shedding load inside a network path has been widely studied in literature~\cite{sigcomm14:Chowdhury, sigcomm14:Dogar, socc12:He}, this paper more focuses on how to build a practical solution of overload control for an operational microservice system. 
The main contributions of this paper are to (1) present the design of \dagor{}, (2) share experiences of operating overload control in the \wechat{} business system, and (3) demonstrate the capability of \dagor{} through experimental evaluation.

The rest of the paper is organized as follows. 
\autoref{sec:background} introduces the overall service architecture of \wechat{} backend as well as workload dynamics that it usually faces. 
\autoref{sec:scenario} describes the overload scenarios under \wechat{}'s microservice architecture.
\autoref{sec:design} presents the design of \dagor{} overload control and its adoption in \wechat{}.
We conduct experiments in \autoref{sec:eval} to evaluate \dagor{}, review related work in \autoref{sec:related_work}, and finally conclude the paper in \autoref{sec:concl}.

\section{Background} 
\label{sec:background}

As a background, we introduce the service architecture of \wechat{} backend which is supporting more than $3000$ mobile services, including instant messaging, social networking, mobile payment and third-party authorization.

\subsection{Service Architecture of \wechat{}}
\label{subsec:service_arch}

The \wechat{} backend is constructed based on the microservice architecture, in which common services recursively compose into complex services with a wide range of functionality.
The interrelation among different services in \wechat{} can be modeled as a directed acyclic graph (DAG), where a vertex represents a distinct service and an edge indicates the call path between two services.
Specifically, we classify services into two categories: \textit{basic service} and \textit{leap service}.
The out-degree (i.e., number of outbound edges) of a basic service in the service DAG is zero, whereas that of a leap service is non-zero. 
In other words, a leap service can invoke other services, either basic or leap, to trigger a downstream service task, and any task is eventually sunk to a basic service which will not further invoke any other services. 
Specially, a leap service with in-degree (i.e., number of inbound edges) being zero in the service DAG is referred as an \textit{entry service}. 
The processing of any service request raised by user always starts with an entry service and ends with a basic service.

\begin{figure}[!t]
  \centering
  \includegraphics[width=.82\linewidth]{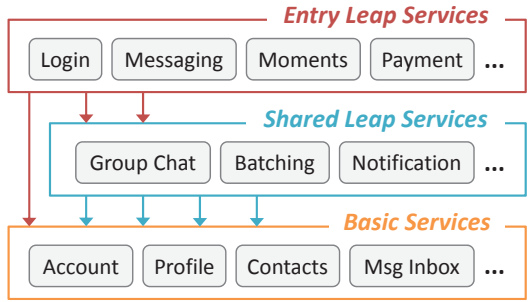}
  \caption{\wechat{}'s microservice architecture.}
  \label{fig:wechat_soa}
\end{figure}

\autoref{fig:wechat_soa} demonstrates the microservice architecture of \wechat{} backend which is hierarchized into three layers.
Hundreds of entry services stay at the top layer.
All the basic services are placed at the bottom layer. 
The middle layer contains all the leap services other than the entry ones.
Every service task is constructed and processed by going through the microservice architecture in a top-down manner. 
In particular, all the basic services are shared among all the leap services for invocation, and they are the end services that serve the user-level purposes\footnote{%
For example, the Account service maintains users' login names and passwords, the Profile service maintains users' nicknames and other personal information, the Contact service maintains a list of friends connected to the user, and the Message Inbox service caches users' incoming and outgoing messages.}. 
Moreover, a leap service in the middle layer is shared by all the entry services as well as other leap services.
Most of the \wechat{} services belong to this layer.

For \wechat{}, the total amount of requests to the entry services is normally $10^{10} \sim 10^{11}$ on a daily basis. 
Each entry service request subsequently triggers more requests to the collaborative microservices to actualize the user-intended behavior.
As a consequence, the \wechat{} backend essentially needs to handle hundreds of millions of service requests per second, and system processing data at such scale is obviously challenging.

\subsection{Deployment of \wechat{} Services}

\wechat{}'s microservice system accommodates more than \todo{$3000$} services running on over \todo{$20000$} machines in the \wechat{} business system, and these numbers keep increasing as \wechat{} is becoming immensely popular.
The microservice architecture allows different development teams to deploy and update their developed services independently. 
As \wechat{} is ever actively evolving, its microservice system has been undergoing fast iteration of service updates.
For instance, \todo{from March to May in 2018}, \wechat{}'s microservice system experienced almost a thousand changes per day on average.
According to our experience of maintaining the \wechat{} backend, any centralized or SLA-based overload control mechanism can hardly afford to support such rapid service changes at large scale.

\subsection{Dynamic Workload} 
\label{subsec:dynamic_workload}

Workload handled by the \wechat{} backend is always varying over time, and the fluctuation pattern differs among diverse situations. 
Typically, the request amount during peak hours is about $3$ times larger than the daily average.
In occasional cases, such as during the period of Chinese Lunar New Year, the peak amount of workload can rise up to around $10$ times of the daily average.
It is challenging to handle such dynamic workload with a wide gap of service request amount. 
Although over-provisioning physical machines can afford such huge workload fluctuation, the solution is obviously uneconomic. 
Instead, it is advisable and more practical by carefully designing the overload control mechanism to adaptively tolerate the workload fluctuation at system runtime. 

\section{Overload in WeChat} 
\label{sec:scenario}

\begin{figure}[!t]
  \centering
  \includegraphics[width=.98\linewidth]{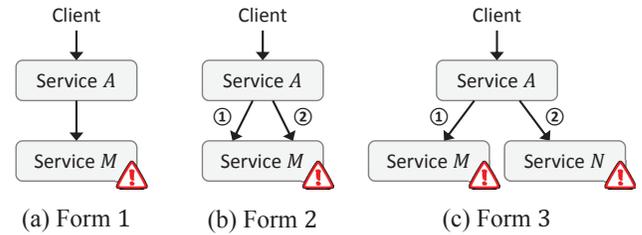}
  \vspace{-1ex}
  \caption{Common overload scenarios.}
  \label{fig:scenario}
\end{figure}

System overload in the microservice architecture can result from various causes. 
The most common ones include load surge, server capacity degradation due to the change of service agreement, network outage, changes of system configuration, software bugs and hardware failures. 
A typical \textit{overload scenario} involves the overloaded services and the service requests along the associated call path. 
In this section, we describe three basic forms of service overload that are complete and able to be used for composing other complex forms of service overload.
The three basic forms are illustrated in \autoref{fig:scenario}, in which the overloaded services are labeled by the attention sign and the associated requests along the call path are denoted by arrows.

\subsection{Overload Scenarios} 
\label{subsec:scenarios}

In Form~1, as shown in \autoref{fig:scenario}.a, overload occurs at service~$M$.
In the \wechat{} business system, service~$M$ usually turns out to be a basic service. 
This is because basic services represent the sink nodes in the service DAG of the microservice architecture, and therefore they are the most active services. 
In \autoref{fig:scenario}.a, the arrows indicate a certain type of request that invokes service~$M$ through service~$A$. 
When service~$M$ is overloaded, all the requests sending to service~$M$ get affected, resulting in delayed response or even request timeout.
Even worse, upstream services (e.g., service~$A$) of the overloaded service are also affected, since they are pending on the responses from the overloaded service. 
This leads to backward propagation of overload from the overloaded service to its upstream services. 

While Form~1 is common in SOA, Form~2 and Form~3 are unique to the large-scale microservice architecture\footnote{%
Form~2 and Form~3 are in fact also common in GFS-like systems~\cite{sosp03:Ghemawat}, where a big file is split into many chunks distributed over different storage servers, and all the chunks need to be retrieved to reconstruct the original file.}.
Form~2, as shown in \autoref{fig:scenario}.b, is similar to Form~1 but involves more than one invocation from service~$A$ to service~$M$. 
Such multiple invocation may be required according to the business logic. 
For example, in an encryption/decryption application, service~$A$ may firstly invoke service~$M$ to decrypt some data, then manipulate the plain-text data locally, and finally invoke service~$M$ again to encrypt the resulting data. 
We term the corresponding overload scenario as \textit{subsequent overload}, which is formally defined as follows. 

\begin{defn}[\textbf{Subsequent Overload}]
Subsequent overload refers to the overload scenario such that there exist more than one overloaded services or the single overloaded service is invoked multiple times by the associated upstream services. 
\end{defn}

In the scenario of subsequent overload, a service task instantiated by the upstream service succeeds if and only if all its issued requests get successful responses. 
Otherwise, processing of a service task is considered failed if any of the service requests sent to the overloaded service is not satisfied. 
Obviously, both Form~2 (in \autoref{fig:scenario}.b) and Form~3 (in \autoref{fig:scenario}.c) belong to subsequent overload.
Subsequent overload in Form~2 is due to the consecutive invocations to the single overloaded service, whereas subsequent overload in Form~3 is caused by the separate invocations to different overloaded services. 

Subsequent overload raises challenges for effective overload control. 
Intuitively, performing load shedding at random when a service becomes overloaded can sustain the system with a saturated throughput. 
However, subsequent overload may greatly degrade system throughput out of anticipation.
This is due to the service constraint of consecutive success of invoking the overloaded service. 
For example, in Form~2 shown in \autoref{fig:scenario}.b, suppose service~$A$ invokes service~$M$ twice, and service~$M$ is configured with capacity $\mathcal{C}$ and performs load shedding at random when it is overloaded. 
We further suppose the current workload feed rate to service~$M$ is $2\mathcal{C}$, and service~$M$ is only called by service~$A$. 
As a consequence, service~$M$ is expected to reject half of the incoming requests, inferring the success rate of each service~$M$ invocation to be $50\%$. 
From the perspective of service~$A$, its success rate regarding the invocations to service~$M$ is $50\% \times 50\% = 25\%$. 
In other words, with $\mathcal{C}$ service~$A$ tasks issued, $2\mathcal{C}$ requests are sent to service~$M$ and only $0.25\mathcal{C}$ service~$A$ tasks eventually survive. 
In contrast, if the workload of service~$A$ is $0.5\mathcal{C}$, then service~$M$ is just saturated without overload and thus the amount of successful service~$A$ tasks is $0.5\mathcal{C}$. 
As can be seen, subsequent overload can lead to the success rate of service~$A$ become very low if each service~$A$ task has to invoke service~$M$ many times.
The same argument also applies to Form~3 shown in \autoref{fig:scenario}.c, with the only difference that the success rate of service~$A$ relies on the product of request success rates of service~$M$ and service $N$.

Among the above three basic forms of service overload, Form~1 and Form~2 dominate the overload cases in the \wechat{} business system, whereas Form~3 appears to be relatively rare.
Towards effective overload control, we emphasize that subsequent overload must be properly handled to sustain the system throughput when the runtime workload is heavy.
Otherwise, simply adopting random load shedding could lead to extremely low (e.g., close to zero) request success rate at the client side when the requesting services are overloaded. 
Such ``service hang'' has been observed in our previous experience of maintaining the \wechat{} backend, and it motivated us to investigate the design of overload control that fits \wechat{}'s microservice architecture.

\subsection{Challenges of Overload Control at Scale} 
\label{subsec:challenges}

Comparing with traditional overload control for the web-oriented three-tier architecture and SOA, overload control for \wechat{}'s microservice architecture has two unique challenges.

First, there is no single entry point for service requests sent to the \wechat{} backend.
This invalidates the conventional approach of centralized load monitoring at a global entry point (e.g., the gateway).
Moreover, a request may invoke many services through a complex call path. 
Even for the same type of requests, the corresponding call paths could be quite different, depending on the request-specific data and service parameters. 
As a consequence, when a particular service becomes overloaded, it is impossible to precisely determine what kind of requests should be dropped in order to mitigate the overload situation. 

Second, even with decentralized overload control in a distributed environment, excessive request aborts could not only waste the computational resources but also affect user experience due to the high latency of service response. 
Especially, the situation becomes severe when subsequent overload happens. 
This calls for some kind of coordination to manage load shedding properly, regarding the request type, priority, call path and service properties. 
However, given the service DAG of the microservice architecture being extremely complex and continuously evolving on the fly, the maintenance cost as well as system overhead for effective coordination with respect to the overloaded services are considered too expensive. 

\section{DAGOR Overload Control}
\label{sec:design}

The overload control scheme designed for \wechat{}'s microservice architecture is called \dagor{}. 
Its design aims to meet the following requirements.

\begin{itemize}
\item \textbf{Service Agnostic}. 
\dagor{} needs to be applicable to all kinds of services in \wechat{}'s microservice architecture, including internal and external third-party services.
To this end, \dagor{} should not rely on any service-specific information to perform overload control. 
Such design consideration of being service agnostic has two advantages. 
First, the overload control mechanism can be highly scalable to support large amount of services in the system, and meanwhile adapt to the dynamics of service deployment. 
This is essential for the ever evolving \wechat{} business, as diverse services deployed in the \wechat{} business system are frequently updated, e.g., new services going online, upgrading existing services and adjusting service configurations.
Second, the semantics of overload control can be decoupled from the business logic of services. 
As a consequence, improper configuration of service does not affect the effectiveness of overload control. 
Conversely, overload detection can help find the implicit flaw of service configuration which causes service overload at runtime. 
This not only benefits service development and debugging/tuning, but also improves system availability as well as robustness.

\item \textbf{Independent but Collaborative}.
In the microservice architecture, a service is usually deployed over a set of physical machines in order to achieve scalability and fault tolerance. 
In practice, workload distribution over the machines is hardly balanced, and the load status of each machine may fluctuate dramatically and frequently, with few common patterns shared among different machines. 
Therefore, overload control should run on the granule of individual machine rather than at the global scale. 
On the other hand, collaborative inter-machine overload control is also considered necessary for handling subsequent overload. 
The collaboration between different machines needs to be service-oriented so that the success rate of the upstream service can match the response rate of the overloaded service, in spite of the occurrence of subsequent overload. 

\item \textbf{Efficient and Fair}.
\dagor{} should be able to sustain the best-effort success rate of service when load shedding becomes inevitable due to overload. 
This infers that the computational resources (i.e., CPU and I/O) wasted on the failed service tasks are minimized. 
Note that those immediately aborted tasks cost little computation and consequentially yield the resource for other useful processing; in contrast, tasks that are partially processed but eventually get aborted waste the computation spent on them.
Therefore, the efficiency of overload control refers to how the mechanism can minimize the waste of computational resources spent on the partial processing of service tasks.
Moreover, when a service gets overloaded, its upstream services should be able to sustain roughly the same saturated throughput despite how many invocations an upstream service task makes to the overloaded service. 
This reflects the fairness of the overload control. 
\end{itemize}

\subsection{Overload Detection}
\label{subsec:overload_detect}

\dagor{} adopts decentralized overload control at the server granule and thus each server monitors its load status to detect potential overload in time. 
For load monitoring towards overload detection in a single server, a wide range of performance metrics have been studied in literature, including throughput, latency, CPU utilization, packet rate, number of pending requests, request processing time, etc.~\cite{queue12:Nichols, queue15:Maurer, usits03:Welsh}. 
\dagor{} by design uses the average waiting time of requests in the pending queue (or \textit{queuing time} for short) to to profile the load status of a server.
The queuing time of a request is measured by the time difference between the request arrival and its processing being started at the server.

The rationale of monitoring the queuing time for overload detection is not so obvious at the first sight. 
One immediate question is: Why not consider using the response time\footnote{%
Response time is defined as the time difference between the request arriving at the server and the corresponding response sent out from the server.} 
instead? 
We argue that the queuing time can be more accurate to reflect the load status than the response time. 
Comparing with the queuing time, the response time additionally counts the request processing time.
In particular, the time for processing a basic service request is purely determined by the local processing, whereas the processing time for a leap service request further involves the cost of processing the downstream service requests. 
This results in the measurement of response time being recursive along the service call path, making the metric failed to individually reflect the load status of a service or a server. 
In contrast, the queuing time is only affected by the capability of local processing of a server.
When a server becomes overloaded due to resource exhaustion, the queuing time rises proportional to the excess workload.
On the other hand, the queuing time would stay at a low level if the server has abundant resources to consume the incoming requests.
Even if the downstream server may respond slowly, queuing time of the upstream server is not affected as long as it has sufficient resource to accommodate the pending service tasks, though its response time does rise according to the slow response of the downstream services. 
In other words, the response time of a server increases whenever the response time of its downstream servers increases, even though the server itself is not overloaded. 
This provides a strong evidence that the queuing time can reflect the actual load status of a server, whereas the response time is prone to false positives of overload. 

Another question is: Why is not CPU utilization used as an overload indicator?
It is true that high CPU utilization in a server indicates that the server is handling high load. 
However, high load does not necessarily infer overload. 
As long as the server can handle requests in a timely manner (e.g., as reflected by the low queuing time), it is not considered to be overloaded, even if its CPU utilization stays high.

Load monitoring of \dagor{} is window-based, and the window constraint is compounded of a fixed time interval and a maximum number of requests within the time interval. 
In the \wechat{} business system, each server refreshes its monitoring status of the average request queuing time every second or every $2000$ requests, whenever either criteria is met.
Such compounded constraint ensures that the monitoring can immediately catch up with the load changes in spite of the workload dynamics. 
For overload detection, given the default timeout of each service task being $500$ ms in \wechat{}, the threshold of the average request queuing time to indicate server overload is set to $20$ ms.
Such empirical configurations have been applied in the \wechat{} business system for more than \todo{five} years with its effectiveness proven by the system robustness with respect to \wechat{} business activities.

\subsection{Service Admission Control}
\label{subsec:admission_control}

Once overload is detected, the corresponding overload control is based on service admission control.
\dagor{} contains a bundle of service admission control strategies. 
We first introduce two types of the priority-based admission control adopted in \dagor{} overload control, and then extend the technique to further support adaptive and collaborative admission control.

\subsubsection{Business-oriented Admission Control}
\label{subsec:business_priority}

\wechat{} services are internally prioritized based on their business significance and impact on user experience, so are the corresponding service requests. 
For example, the Login request is of the highest priority, because user cannot interact with other services until he/she completes a successful login. 
Another example is that the \wechatpay{} request has higher priority than the Instant Messaging request. 
This is because users tend to be sensitive to their money-related interactions such as mobile payment, while they are usually able to accept a certain degree of delay or inconsistency in the messaging service.
The operation log of \wechat{} shows that when \wechatpay{} and Instant Messaging experience a similar period of service unavailability, user's complaint against the \wechatpay{} service is $100$ times more than that against the Instant Messaging service. 
Similar situation also applies to Instant Messaging versus \moments{}, as user expects more timely delivery of content with Instant Messaging than with \moments{}. 

\begin{figure}[!t]
  \centering
  \includegraphics[width=.8\linewidth]{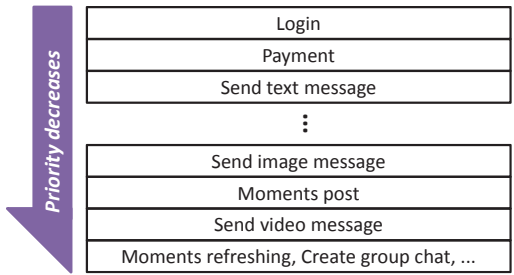}
  \caption{Hash table storing the business priorities of actions to perform in the \wechat{} entry services.}
  \label{fig:business_priority}
\end{figure}

Business-oriented admission control in \dagor{} is to assign a business priority to each user request and enforce all its subsequent requests inherit the same business priority. 
When a service becomes overloaded, its load shedding routine will give priority to discarding low-priority requests, yielding resources for high-priority requests. 
The business priority of a user request as well as its subsequent requests along the call path is determined by the type of action to perform in the entry service.
As there exist hundreds of entry services in \wechat{}'s microservice architecture, the number of different actions to perform in the entry services is of hundreds. 
The business priorities are predefined and stored in a hash table, which is replicated to all the \wechat{} backend servers that host the entry services. 
An item in the hash table records the mapping from an action ID (representing a distinct type of action) to a priority value.
\autoref{fig:business_priority} illustrates the logical structure of the action-priority hash table. 
Note that the hash table does not contain all types of actions.
By default, action types that are missing in the hash table correspond to the lowest priority.
Only those intentionally prioritized action types are recorded in the hash table, \todo{with smaller priority value indicating higher priority of the action}.
This results in the hash table containing only a few tens of entries. 
Since the set of prioritized actions to perform in the entry service are empirically stable, the hash table remains compact with rare changes despite the rapid evolution of \wechat{} business. 

Whenever a service request triggers a subsequent request to the downstream service, the business priority value is copied to the downstream request. 
By recursion, service requests belonging to the same call path share an identical business priority. 
This is based on the presumption that the success of any service request relies on the conjunctive success of all its subsequent requests to the downstream services. 
As the business priority is independent to the business logic of any service, \dagor{}'s service admission control based on the business priority is service agnostic. 
Moreover, the above business-oriented admission control is easy to maintain, especially for the complex and highly dynamic microservice architecture such as the \wechat{} backend. 
On the one hand, the assignment of business priority is done in the entry services by referring to the action-priority hash table, which is seldom changed over time\footnote{\todo{%
The action-priority hash table may be modified on the fly for performance tuning or ad-hoc service support. 
But this happens very rarely in the \wechat{} business system, e.g., once or twice per year.}}. 
This makes the strategy of business priority assignment relatively stable. 
On the other hand, the dynamics of \wechat{}'s microservice architecture are generally reflected in the changes of basic services and leap services other than the entry services. 
Since the business priorities of requests to these frequently changing services are inherited from the upstream service requests, developers of these services can simply apply the functionality of business-oriented admission control as a black box without concerning the setting of business priority\footnote{%
We used to additionally provide APIs for service developer to adjust the business priority of request specifically for the service. 
However, the solution turned out to be not only extremely difficult to manage among different development teams, but also error-prone with respect to the overload control. 
Consequently, we deprecated the APIs for setting service-specific business priority in the \wechat{} business system.}.

\subsubsection{User-oriented Admission Control}

The aforementioned strategy of business-oriented admission control constrains the decision of dropping a request to be determined by the business priority of the request. 
In other words, for load shedding upon service overload, the business-oriented admission control presumes requests with the same business priority are either all discarded or all consumed by the service. 
However, partially discarding requests with respect to the same business priority in an overloaded service is sometimes inevitable. 
\todo{%
Such inevitability emerges when the admission level of business priority of the overloaded service is fluctuating around its ``ideal optimality''. 
To elaborate, let us consider the following scenario where load shedding in an overloaded service is solely based on the business-oriented admission control. 
Suppose the current admission level of business priority is $\tau$ but the service is still overloaded. 
Then the admission level is adjusted to $\tau - 1$, i.e., all requests with business priority value greater than or equal to $\tau$ are discarded by the service. 
However, system soon detects that the service is underloaded with such admission level. 
As a consequence, the admission level is set back to $\tau$, and then the service quickly becomes overloaded again.
The above scenario continues to repeat.
As a result, the related requests with business priority equal to $\tau$ are in fact partially discarded by the service in the above scenario.}

Partially discarding requests with the same business priority could bring on issue caused by subsequent overload, because these requests are actually discarded at random under such situation.
To tackle this issue, we propose the user-oriented admission control as a compensation for the business-oriented admission control. 
User-oriented admission control in \dagor{} is based on the user priority. 
\todo{%
The user priority is dynamically generated by the entry service through a hash function that takes the user ID as an argument. 
Each entry service changes its hash function every hour. 
As a consequence, requests from the same user are likely to be assigned to the same user priority within one hour, but different user priorities across hours.}
The rationality for the above strategy of user priority generation is twofold.
On the one hand, the one-hour period of hash function alternation allows user to obtain a relatively consistent quality of service for a long period of time.
On the other hand, the alternation of hash function takes into account the fairness among users, as high priorities are granted to different users over hours of the day. 
Like the business priority, the user priority is also bound to all the service requests belonging to the same call path. 

The strategy of user-oriented admission control cooperates with the business-oriented admission control. 
\todo{%
For requests with business priority equal to the admission level of business priority of the overloaded service, the corresponding load shedding operation gives priority to the ones with high user priority.}
By doing so, once a request from service~$A$ to the overloaded service~$M$ gets a successful response, the subsequent request from service~$A$ to service~$M$ is very likely to also get a successful response. 
This resolves the issue caused by subsequent overload of Form~2 as shown in \autoref{fig:scenario}.b.
Moreover, subsequent overload of Form~3 as shown in \autoref{fig:scenario}.c can also be properly handled in a similar way.
Suppose the upstream service~$A$ invokes two overloaded dependent services, i.e., service~$M$ and service $N$, in order. 
If the admission levels regarding the business and user priorities of service~$M$ are more restricted than that of service $N$, then subsequent overload can be eliminated in Form~3.
This is because a request being admitted by service~$M$ implies the admission of the subsequent request to service $N$, due to the relaxed admission levels.
Such condition of admission levels between the dependent services in Form~3 usually holds in the \wechat{} business system, as the preceding service is prone to more severe overload.

\begin{figure}[!t]
  \centering
  \includegraphics[width=.98\linewidth]{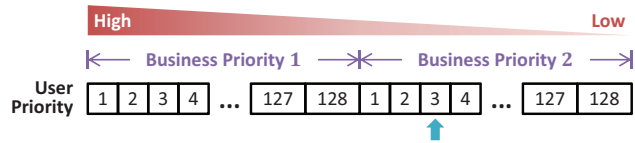}
  \vspace{-1.5ex}
  \caption{The compound admission level.}
  \label{fig:compound_admission}
\end{figure}

\paragraph{Session-oriented Admission Control}
Other than the user-oriented admission control, we have ever proposed the session-oriented admission control to address the same issue caused by solely applying the business-oriented admission control. 
The session-oriented admission control is based on the \textit{session priority}, whose generation as well as functionality are similar to that of the user priority as described before. 
The only difference is that the hash function of generating the session priority alternatively takes the session ID as an argument. 
A session ID is assigned to a user upon his/her successful login, which indicates the start of a user session. 
A session generally ends with a confirmed logout performed by the user. 
When the same user performs another login after his/her prior logout, another session is created with a different session ID and thus a new session priority is generated accordingly, even though the hash function remains unchanged. 
In terms of overload control in \dagor{}, the session-oriented admission control is as effective as the user-oriented admission control. 
But our operational experience with the \wechat{} business system shows that the session-oriented admission control tends to degrade user experience. 
This is due to the natural user behavior where \wechat{} users often prefer to logout and immediately login again whenever they encounter service unavailability in the app. 
The same phenomenon also frequently arises in other mobile apps. 
Through the logout and immediate login, user obtains a refreshed session ID. 
As a consequence, the session-oriented admission control assigns the user a new session priority, which could be high enough to grant his/her service requests in the overloaded service backend. 
Gradually, the user may figure out the ``trick'' that enables him/her to escape from service unavailability via re-login. 
When a trick is repeatedly validated to be effective, it tends to become a user habit. 
However, such trick does not help mitigate the actual service overload occurring at the system backend. 
Moreover, it would introduce extra user requests due to the misleading logout and login, further deteriorating the overload situation and thus eventually affecting the user experience of the majority of users. 
In contrast, user's instant re-login does not affect his/her user priority in the user-oriented admission control. 
Hence, we prefer the user-oriented admission control over the session-oriented one in \dagor{} overload control.

\subsubsection{Adaptive Admission Control}
\label{subsec:adaptive_ac}

Load status in the microservice system is always dynamically changing. 
A service becomes sensitive to the change of its load status when it is overloaded, since the corresponding load shedding strategy is dependent on the volume of the excess workload.
Therefore, the priority-based admission levels should be able to adapt to the load status towards effective load shedding with minimized impact on the quality of the overall service.
When the overload situation is severe, the admission levels should be restricted to reject more incoming requests; on the contrary, the admission levels should be relaxed when the overload situation becomes mild. 
In the complex microservice architecture, the adjustment of admission levels of the overloaded services needs to be automatic and adaptive. 
This calls for the necessity of adaptive admission control in the overload control of \dagor{}.

\dagor{} adjusts the admission levels of the overloaded services with respect to their individual load status.
As illustrated in \autoref{fig:compound_admission}, \dagor{} uses the \textit{compound admission level} which is composed of the business and user priorities. 
Each admission level of business priority is attached with $128$ admission levels of user priority.
Let $\vB$ and $\vU$ denote the business priority and the user priority respectively, and the compound admission level is denoted by $(\vB,\ \vU)$. 
A cursor, denoted by an arrow in \autoref{fig:compound_admission}, indicates the current admission level\footnote{%
If not specified, we refer the admission level to the compound admission level in the rest of the paper} 
to be $(2,\ 3)$, which is interpreted as all the requests with $\vB > 2$ and the requests with $\vB = 2$ but $\vU > 3$ will be shed by the overloaded service. 
Moving the cursor leftwards represents raising the admission level, i.e., restricting the business and user priorities. 

As mentioned in \autoref{subsec:business_priority}, \dagor{} maintains tens of distinct admission levels of business priority. 
With each admission level of business priority attached with $128$ admission levels of user priority, the resulting amount of the compound admission levels is tens of thousands. 
Adjustment of the compound admission level is at the granule of user priority. 
To search for the appropriate admission level according to the load status, a naive approach is by trying each admission level one by one starting from the lowest, i.e., moving the cursor from the far right to the left in \autoref{fig:compound_admission}.
\todo{%
Note that for each setting of the admission level, server has to take a while to validate its effectiveness, since the load status is aggregated within a certain time interval.
As the incoming requests are unlikely to be distributed evenly over the admission levels, such naive approach tends to be awkward to find the right setting. 
This is because every adjustment of admission level in the naive approach, i.e., moving the cursor leftwards by one user priority, exerts marginal impact on the overload situation but takes time to validate its sufficiency.
Therefore, by scanning the admission levels in the naive way, the adjustment of admission level can hardly meet the real-time requirement for the adaptive admission control.}
An immediate improvement based on the above naive approach is to perform binary search instead of linear scan. 
This makes the search complexity reduces from $\mathcal{O}(n)$ to $\mathcal{O}(\log n)$ where $n$ represents the number of compound admission levels in total. 
Regarding the actual amount of admission levels in \wechat{} is at the scale of $10^{4}$, the $\mathcal{O}(\log n)$ search involves about a dozen trials of the adjustment of admission level, which is still considered away from efficiency. 

\begin{algorithm}[!t]
\AlgoFontSize
\DontPrintSemicolon

\KwGlobal{Range of admission levels of business priority $\vBmin \dots \vBmax$}
\KwGlobal{Range of admission levels of user priority $\vUmin \dots \vUmax$}
\KwGlobal{Incoming request counters $C[\vBmin \dots \vBmax][\vUmin \dots \vUmax]$}
\KwGlobal{Incoming request counter $\vN$}
\KwGlobal{Admitted request counter $\vNadm$}
\KwGlobal{Compound admission level $(\vBopt,\ \vUopt)$}

\BlankLine

\Proc(\tcc*[f]{at the beginning of period}){\fResetHistogram{}}{
  $\vN, \vNadm \gets 0$\;
  \lForEach{$c \in C$}{$c \gets 0$}
}

\BlankLine

\KwIn{Service request $r$}
\Proc{\fUpdateHistogram{$r$}}{
  $N \gets N + 1$\;
  $C[r.\vB][r.\vU] \gets C[r.\vB][r.\vU] + 1$\;
  \If{$r.\vB < \vBopt$ \Or $($ $r.\vB = \vBopt$ \And $r.\vU \leq \vUopt$ $)$}{
    $\vNadm \gets \vNadm + 1$
  }
}

\BlankLine

\KwIn{Boolean flag $\vFoverload$ indicating overload}
\Proc(\tcc*[f]{at the end of period}){\fUpdateAdmitLevel{$\vFoverload$}}{
  $\vNprefix \gets \vNadm$\;
  \uIf{$\vFoverload = \cTrue$}{
    $\vNexp \gets (1 - \alpha) \cdot \vNadm$\;
    \While{$\vNprefix > \vNexp$ \And $(\vBopt,\ \vUopt) > (\vBmin,\ \vUmin)$}{
      $\vUopt \gets \vUopt - 1$\;
      \If{$\vUopt < \vUmin$}{
        $\vBopt \gets \vBopt - 1$\;
        $\vUopt \gets \vUmax$\;
      }
      $\vNprefix \gets \vNprefix - C[\vBopt][\vUopt]$\;
    }
  }\Else{
    $\vNexp \gets \vNadm + \beta \cdot \vN$\;
    \While{$\vNprefix < \vNexp$ \And $(\vBopt,\ \vUopt) < (\vBmax,\ \vUmax)$}{
      $\vUopt \gets \vUopt + 1$\;
      \If{$\vUopt > \vUmax$}{
        $\vBopt \gets \vBopt + 1$\;
        $\vUopt \gets \vUmin$\;
      }
      $\vNprefix \gets \vNprefix + C[\vBopt][\vUopt]$\;
    }
  }
}

%

\caption{Adaptive admission control.}
\label{algo:adaptive_ac}
\end{algorithm}

Towards adaptive admission control with efficiency, \dagor{} exploits a histogram of requests to quickly figure out the appropriate setting of admission level with respect to the load status.
The histogram helps to reveal the approximate distribution of requests over the admission priorities. 
In particular, each server maintains an array of counters, each of which corresponds to a compound admission level indexed by $(\vB,\ \vU)$. 
Each counter counts the number of the incoming requests associated with the corresponding business and user priorities. 
\dagor{} periodically adjusts the admission level of load shedding as well as resets the counters, and the period is consistent to the window size for overload detection as described in \autoref{subsec:overload_detect}, e.g., every second or every $2000$ requests in the \wechat{} business system.
For each period, if an overload is detected, the server sets the expected amount of admitting requests in the next period to be $\alpha$ (in percentage) less than that in the current period; otherwise, the expectation of request amount in the subsequent period is increased by $\beta$ (in percentage) of the incoming requests in the current period.
Empirically, we set $\alpha = 5\%$ and $\beta = 1\%$ in the \wechat{} business system.
Given the expected amount of admitting requests, the admission level is calculated with respect to the maximum prefix sum in the histogram near that amount, and is adjusted based on its current state and whether or not an overload is detected.
Let $\vBopt$ and $\vUopt$ be the optimal settings of the admission levels of business priority and user priority respectively. 
The optimal setting of the compound admission level $(\vBopt,\ \vUopt)$ is determined by the constraint such that the sum of counters with $(\vB,\ \vU) \leq (\vBopt,\ \vUopt)$\footnote{%
For two admission levels $(\vB_{1},\ \vU_{1})$ and $(\vB_{2},\ \vU_{2})$, we have $(\vB_{1},\ \vU_{1}) < (\vB_{2},\ \vU_{2})$ if $\vB_{1} < \vB_{2}$, or $\vB_{1} = \vB_{2}$ but $\vU_{1} < \vU_{2}$.} 
is just below (resp.\ exceeding) the expected amount when overload is indicated (resp.\ suppressed). 
\autoref{algo:adaptive_ac} summaries the procedures of adaptive admission control in \dagor{}.
Note that even if the system is not overloaded at some point, some requests may still be discarded due to the current settings of admission levels, especially when the system is undergoing recovery from a previous overload. 

Obviously, the above approach only involves a single trial of validation per adjustment of the admission level. 
Hence, it is much more efficient than the aforementioned naive approaches, and can satisfy the real-time requirement for the adaptive admission control. 

\subsubsection{Collaborative Admission Control}
\label{subsec:collaborative_ac}

\dagor{} enforces the overloaded server to perform load shedding based on the priority-based admission control. 
Regarding message passing, a request that is destined to be shed by the overloaded downstream server still has to be sent from the upstream server, and the downstream server subsequently sends the corresponding response back to the upstream server to inform the rejection of the request.
Such round-trip of message passing for unsuccessful request processing not only wastes the network bandwidth but also consumes the tense resource of the overloaded server, e.g., for serializing/deserializing network messages. 
To save network bandwidth and reduce the burden on the overloaded server to handle excessive requests, it is advised to reject the requests that are destined to be shed early at the upstream server. 
To this end, \dagor{} enables collaborative admission control between the overloaded server and its upstream servers. 
In particular, a server piggybacks its current admission level $(\vB,\ \vU)$ to each response message that it sends to the upstream server. 
When the upstream server receives the response, it learns the latest admission level of the downstream server. 
By doing so, whenever the upstream server intends to send request to the downstream server, it performs a local admission control on the request according to the stored admission level of the downstream server. 
As a consequence, requests destined to be rejected by the downstream server tends to be shed early at the upstream server, and requests actually sent out from the upstream server tends to be admitted by the downstream server. 
Therefore, while the strategy of admission control in a server is independently determined by the server itself, the actual load shedding is performed by its related upstream servers. 
Such collaboration between the upstream and downstream servers greatly benefits the improved efficiency of overload control in the microservice architecture.

\begin{figure}[!t]
  \centering
  \includegraphics[width=.98\linewidth]{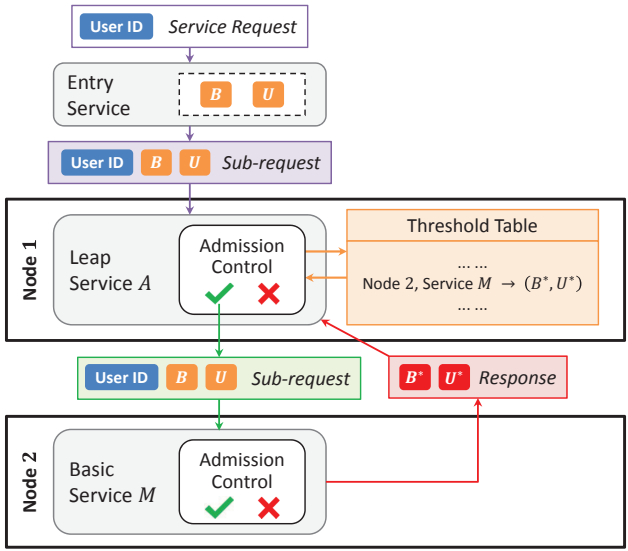}
  \caption{Workflow of \dagor{} overload control.}
  \label{fig:workflow}
\end{figure}

\subsection{Workflow of Overload Control}

Based on the aforementioned strategies of service admission control, we now depict the overall workflow of \dagor{} overload control as illustrated in \autoref{fig:workflow}.

\begin{enumerate}
\item When a user request arrives at the microservice system, it is routed to the related entry service. 
The entry service assigns the business and user priorities to the request, and all the subsequent requests to the downstream services inherit the same priorities which are encapsulated into the request header.

\item Each service invokes one or more downstream services according to the business logic. 
Service requests and responses are delivered through message passing.

\item When a service receives a request, it performs the priority-based admission control based on its current admission level. 
The service periodically adjusts its admission level with respect to the load status. 
When a service intends to send a subsequent request to a downstream service, it performs the local admission control based on the stored admission level of the downstream service. 
The upstream service only sends out the requests that are admitted by the local admission control. 

\item When a downstream service sends a response to an upstream service, it piggybacks its current admission level to the response message. 

\item When the upstream service received the response, it extracts the information of admission level from the message and updates the corresponding local record for the downstream service accordingly. 
\end{enumerate}

\dagor{} satisfies all the requirements of overload control in the microservice architecture as we proposed at the beginning of this section. 
First, \dagor{} is \textit{service agnostic}, since the strategies of service admission control are based on the business and user priorities that are orthogonal to business logic. 
This makes \dagor{} generally applicable to microservice systems. 
Second, service admission control of \dagor{} is \textit{independent but collaborative}, as the admission levels are determined by the individual services and the admission control is collaboratively performed by the related upstream services. 
This makes \dagor{} highly scalable to be adopted for the large-scale, timely evolving microservice architecture.
Third, \dagor{} overload control is \textit{efficient and fair}. 
It can effectively eliminates the performance degradation due to subsequent overload, since all the service requests belonging to the same call path share the identical business and user priorities. 
This makes the upstream service able to sustain its saturated throughput in spite of the overload situation of the downstream service. 


\section{Evaluation}
\label{sec:eval}

\dagor{} has been fully implemented and deployed in the \wechat{} business system for more than \todo{five} years.
It has greatly enhanced the robustness of the \wechat{} service backend and helped \wechat{} survive in various situations of high load operation, including those in daily peak hours as well as the period of special events such as the eve of Chinese Lunar New Year.
In this section, we conduct an experimental study to evaluate the design of \dagor{} and compare its effectiveness with state-of-the-art load management techniques. 

\subsection{Experimental Setup}

All experiments run on an in-house cluster, where each node is equipped with an Intel Xeon \todo{E5-2698} @ $2.3$ GHz CPU and $64$ GB DDR3 memory. 
All nodes in the cluster are connected via $10$ Gigabit Ethernet. 

\textbf{Workloads.}
To evaluate \dagor{} independently, we implement a stress test framework that simulates the encryption service used in the \wechat{} business system. 
Specifically, an encryption service, denoted as $M$, is exclusively deployed over $3$ servers and prone to be overloaded with the saturated throughput being around $750$ queries per second (QPS). 
A simple messaging service, denoted as $A$, is deployed over another $3$ servers to process the predefined service tasks by invoking service~$M$ as many times as requested by the workload.
Workload is synthetically generated by $20$ application servers, which are responsible for generating service tasks and never overloaded in the experiments. 
Each service task is programmed to invoke service~$M$ one or multiple times through service~$A$, and the success of the task is determined by the conjunctive success of those invocations\footnote{%
In case of rejection, the same request of invocation will be resent up to three times.}. 
Let \AMX{} denote the workload consisting of tasks with $x$-invocation to service~$M$. 
Four types of workload, namely \AM{}, \AMM{}, \AMMM{} and \AMMMM{}, are used in the experiments. 
Regarding the overload scenario, \AM{} corresponds to simple overload, while \AMM{}, \AMMM{} and \AMMMM{} correspond to subsequent overload.

\begin{figure*}[!t]
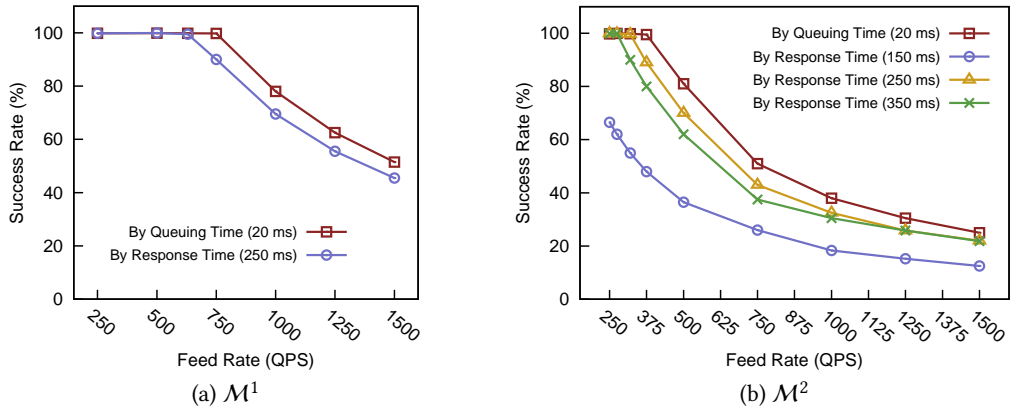

  \centering
  \begin{minipage}[b]{.3339\linewidth} 
    \centering
    \includegraphics[width=\linewidth]{\Exp{eps}{overload-detect-m1}}\\
    (a) \AM{}
  \end{minipage}
  \hspace{2em}
  \begin{minipage}[b]{.3975\linewidth} 
    \centering
    \includegraphics[width=\linewidth]{\Exp{eps}{overload-detect-m2}}\\
    (b) \AMM{}
  \end{minipage}
  \vspace{-0.5ex}
  \caption{Overload detection by different indicators of load profiling: queuing time vs. response time.}
  \label{exp:od}
\end{figure*}

\begin{figure*}[!t]
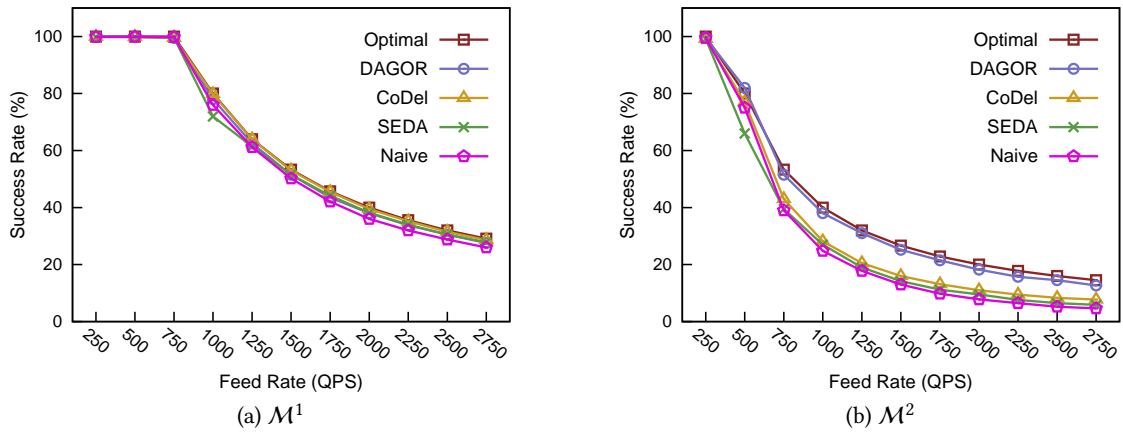

  \centering
  \begin{minipage}[b]{.42\linewidth}
      \centering
      \includegraphics[width=.95\linewidth]{\Exp{eps}{admission-control-m1}} \\
    (a) \AM{}
  \end{minipage}
  \hspace{1.5em}
  \begin{minipage}[b]{.42\linewidth}
      \centering
      \includegraphics[width=.95\linewidth]{\Exp{eps}{admission-control-m2}} \\
    (b) \AMM{}
  \end{minipage}
  \vspace{-0.5ex}
  \caption{Overload control with increasing workload.}
  \label{exp:om_against_input}
\end{figure*}

\subsection{Overload Detection}

We first evaluate \dagor{}'s overload detection, which serves as the entry point of overload control.
%
In particular, we experimentally verify \dagor{}'s design choice of adopting the average request queuing time rather than response time as the indicator of load status for overload detection, as discussed in \autoref{subsec:overload_detect}.
To this end, we additionally implement a \dagor{} variant whose overload detection refers to the average response time of requests over the monitoring window, i.e., every second or every $2000$ requests whenever either is met. 
Let \dagor{}$_q$ (resp.\ \dagor{}$_r$) be the \dagor{} implementations with overload detection based on the request queuing time (resp.\ response time). 
We conduct experiments by running workloads of \AM{} and \AMM{} individually, varying the feed rate from $250$ QPS to $1500$ QPS. 
\autoref{exp:od} shows the comparison between \dagor{}$_q$ and \dagor{}$_r$.
For simple overload as in \autoref{exp:od}.a, we set the thresholds of average queuing time and response time to $20$ ms and $250$ ms respectively. 
As can be seen from the results, \dagor{}$_r$ starts load shedding when the feed rate reaches $630$ QPS, whereas \dagor{}$_q$ can postpone the load shedding until $750$ QPS of the input.
This implies the load profiling based on the response time is prone to false positives of overload. 
\autoref{exp:od}.b further confirms this fact with subsequent overload involved, i.e., by running the \AMM{} workload. 
In addition to the settings as in \autoref{exp:od}.a, we also measure the curves of success rate for \dagor{}$_r$ with the threshold of response time set to $150$ ms and $350$ ms in \autoref{exp:od}.b.
The results show that \dagor{}$_r$ exhibits best performance when the threshold of response time is set to $250$ ms.
However, the optimal configuration of \dagor{}$_r$ is difficult to tune in practice, since the request response time contains the request processing time which is service-specific. 
In contrast, apart from the superior performance of \dagor{}$_q$, its configuration is easy to be fine-tuned because the request queuing time is irrelevant to any service logic.

\subsection{Service Admission Control}

Next, we evaluate \dagor{}'s service admission control for load management. 
As described in \autoref{subsec:admission_control}, the basis of \dagor{}'s service admission control is based on priority, which is further devised to be business-oriented and user-oriented.
The business-oriented admission control is commonly adopted in state-of-the-art load management techniques~\cite{sosp01:Welsh, queue12:Nichols}. 
\dagor{} is novel in its additional use of user-oriented priority for fine-grained load management towards improved end-to-end service quality. 
Moreover, \dagor{} overload control can adaptively adjust service admission levels in a real-time manner through adaptive admission control, and optimize the load shedding runtime with the adoption of collaborative admission control. 
These distinct strategies distinguishes \dagor{} from existing techniques of load management by mechanism. 
Hence, in order to verify the effectiveness of \dagor{}'s innovation in terms of service admission control, we compare \dagor{} with state-of-the-art techniques, i.e., \codel{}~\cite{queue12:Nichols} and \seda{}~\cite{sosp01:Welsh}, through experiments with business priority being fixed for all the generated query requests.
In addition, we also employ a naive approach of service admission control such that the overloaded service~$M$ performs load shedding at random. 
Such naive approach serves as the baseline in the experiments. 

\begin{figure}[!t]
  \centering
  \includegraphics[width=.85\linewidth]{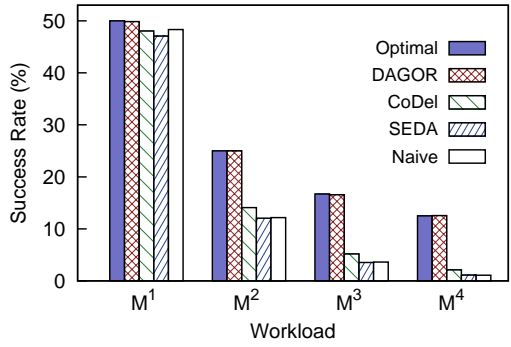}
  \vspace{-0.5ex}
  \caption{Overload control with different types of workload.}
  \label{exp:om_against_type}
\end{figure}

\begin{figure*}[!t]
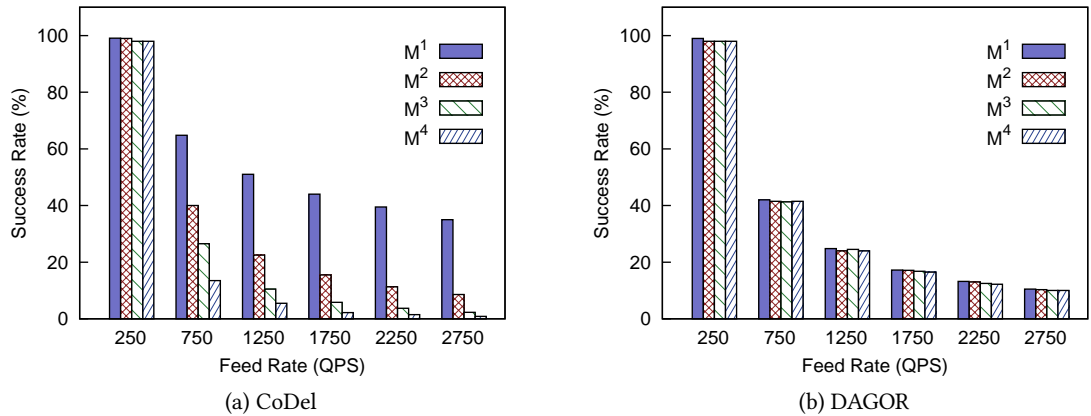

  \centering
  \begin{minipage}[b]{.4\linewidth}
    \centering
    \includegraphics[width=\linewidth]{\Exp{eps}{fair-codel}} \\
    (a) \codel{}
  \end{minipage}
  \hspace{1.5em}
  \begin{minipage}[b]{.4\linewidth}
    \centering
    \includegraphics[width=\linewidth]{\Exp{eps}{fair-dagor}} \\
    (b) \dagor{}
  \end{minipage}
  \vspace{-0.5ex}
  \caption{Fairness of overload control.}
  \label{exp:fairness}
\end{figure*}

\autoref{exp:om_against_input} demonstrates the comparison results by running workloads of 
\AM{} and \AMM{} respectively. 
Each figure compares the success rate of the upstream service requests with the adoption of different service admission control techniques.
In addition, we also depict the theoretically optimal success rate of the upstream service when the corresponding downstream services are overloaded. 
The optimal success rate is calculated by $f_{\textnormal{sat}} / f$, where $f_{\textnormal{sat}}$ is the maximum feed rate that makes the downstream service just saturated, and $f$ refers to the actual feed rate when the downstream service is overloaded.
As can be seen from \autoref{exp:om_against_input}.a, all the overload control techniques perform roughly the same for simple overload (i.e., \AM{}). 
However, when subsequent overload is involved as shown in \autoref{exp:om_against_input}.b, \dagor{} exhibits around $50\%$ higher success rate than \codel{} and \seda{} in the workload of \AMM{}.
\autoref{exp:om_against_type}, measured by fixing the feed rate to $1500$ QPS, further shows the greater advantage of \dagor{} with the increment of subsequent overload in the workloads of \AMMM{} and \AMMMM{}. 
Moreover, the request success rate contributed by \dagor{} is close to the optimal in all the above results.
Such superior of \dagor{} is due to its effective suppression of subsequent overload through the priority-based admission control, especially the user-oriented strategy. 

\subsection{Fairness}

Finally, we evaluate the fairness of overload control with respect to different overload situations. 
The \textit{fairness} refers to whether the overload control mechanism biased towards one or more specific overload forms.  
To that end, we run mixed workload comprising \AM{}, \AMM{}, \AMMM{} and \AMMMM{} requests with uniform distribution.
The requests are issued concurrently by the clients with the feed rate varying from $250$ to $2750$ QPS. 
\todo{The business priority and user priority of each request are chosen at random in a fixed range.}
We compare the fairness of request success rates between \dagor{} and \codel{}, and the results shown in \autoref{exp:fairness}. 
As can be seen, \codel{} favors simple overload (i.e., \AM{}) over subsequent overload (e.g., \AMM{}, \AMMM{} and \AMMMM{}). 
Such bias of \codel{} renders its overload control dependent on the service workflow, where the more complex logic tends more likely to fail when the system is experiencing overload.
In contrast, \dagor{} manifests roughly the same success rate for different types of requests. 
This is because the priority-based admission control of \dagor{} greatly reduces the occurrence of subsequent overload, in spite of the number of upstream invocations to the overloaded downstream services. 


\section{Related Work}
\label{sec:related_work}

Plenty of existing research of overload control has been devoted to real-time databases~\cite{rtcsa99:Hansson, rtds97:Bestavros}, stream processing systems~\cite{sigmod15:Lin, vldb07:Tatbul, vldb06:Xing, icdew06:Tatbul, vldb04:Chandrasekaran} and sensor networks~\cite{tosn07:Wan}. 
For networked services, techniques of overload control have been mainly proposed in the context of web services~\cite{www04:Elnikety}.
However, most of these techniques are designed for the traditional monolithic service architecture and they do not apply to modern large-scale online service systems that are often built based on the SOA.
To the best of our knowledge, \dagor{} is the first to specifically address the issue of overload control for large-scale microservice architectures.
As the microservice architecture is generally deemed to belong to the family of SOA, we closely review the emerging techniques of overload control for SOA, which are built based on either admission control~\cite{usits03:Welsh, infocom02:Chen, tc02:Cherkasova, www01:Chen, socc17:Suresh} or resource scheduling~\cite{osdi99:Banga, tocs97:Mogul, Book01:Menasce, tkde99:Kant, itpro02:Almeida}.

Controlling network flow by capturing complex interactions of modern cloud systems can mitigate the impact of system overload. Varys~\cite{sigcomm14:Chowdhury} employs the coflow and task abstractions to schedule network flows towards reduction of task completion time. But it relies on the prior knowledge of flow sizes and meanwhile assumes the availability of centralized flow control. This renders the method only applicable to limited size of cloud computing infrastructure. Baraat~\cite{sigcomm14:Dogar} alternatively adopts a FIFO-like scheduling policy to get rid of centralized control, but at the expense of performance. Comparing with these coflow-based control, \dagor{} is not only service agnostic but also independent of network flow characteristics. This makes \dagor{} a non-invasive overload control suitable for microservice architecture. 

Cherkasova~et~al.~\cite{tc02:Cherkasova} proposed the session-based admission control, which monitored the performance of web services by counting the completed sessions. 
When the web services become overloaded, the admission control 
rejects requests for creating new sessions. 
Chen~et~al.~\cite{infocom02:Chen} further proposed the QoS-aware session-based admission control by exploiting the dependencies of sessions to improve the service quality during service overload. 
However, these techniques favor long-lived sessions, making them unsuitable for the \wechat{} application which incorporates tremendous short-lived and medium-lived sessions. 
Differently, \dagor{}'s admission control is user-oriented rather than session-oriented, and hence does not bias any session-related property of the service for overload control.
Moreover, the above session-based techniques rely on a centralized module for overload detection and overload control. 
Such centralized module tends to become the system bottleneck, resulting in limited scalability of the system. 
In contrast, \dagor{} by design is decentralized so that it is highly scalable to support the large-scale microservice architecture. 

Welsh~et~al.~\cite{usits03:Welsh, sosp01:Welsh} proposed the technique of staged overload control, which partitioned the web services into stages with respect to the service semantics and performed overload control for each stage independently. 
Each stage is statically allocated with a resource quota for load constraint. 
Although such mechanism shares some similarity with \dagor{}, its prerequisites of service partitioning and static resource allocation render it inapplicable to the complex, dynamic service architecture such as the \wechat{} backend. 
In contrast, \dagor{} is service agnostic, making it flexible and highly adaptable to the continuously evolving microservice architecture. 

Network overload control targeting at reduced reponse time has been well studied~\cite{sigcomm13:Jalaparti}. Our experience of operating \wechat{} business system shows that latency-oriented overload control is hardly effective when used for the large-scale microservice architecture. This motivated us to employ queuing time for overload detection in \dagor{}. Moreover, techniques discussed in the THEMIS system~\cite{sigmod16:Kalyvianaki} inspire us to take fairness into account in the \dagor{} design.  

\section{Conclusion}
\label{sec:concl}

This paper proposed the \dagor{} overload control for the microservice architecture.
\dagor{} by design is service agnostic, independent but collaborative, efficient and fair.
It is lightweight and generally applicable to the large-scale, timely evolving microservice systems, as well as friendly to cross-team agile development. 
We implemented \dagor{} in the \wechat{} service backend and have been running it in the \wechat{} business system for more than \todo{five} years. 
Apart form its effectiveness proved in the \wechat{} practice, we believe \dagor{} and its design principles are also insightful for other microservice systems. 

\textbf{Lessons Learned.}
Having operated \dagor{} as a production service in the \wechat{} business backend for over five years, we share lessons we have learned from our development experience as well as design principles below:

\begin{itemize}
\item Overload control in the large-scale microservice architecture must be decentralized and autonomous in each service, rather than counting on the centralized resource scheduling. 
This is essential for the overload control framework to scale with the ever evolving microservice system. 

\item The algorithmic design of overload control should take into account a variety of feedback mechanisms, rather than relying solely on the open-loop heuristics.
A concrete example is the strategy of collaborative admission control in \dagor{}.

\item An effective design of overload control is always derived from the comprehensive profiling of the processing behavior in the actual workload. 
This is the basis of \dagor{}'s design choices of using the request queuing time for overload detection as well as devising the two-tier priority-based admission control to prevent subsequent overload. 
\end{itemize}

\begin{acks}
We would like to thank the anonymous reviewers for their valuable comments and constructive suggestions that helped improve the paper.
\end{acks}

\balance

\bibliographystyle{ACM-Reference-Format}
\bibliography{references-trim}

\end{document}